\documentstyle[aps,preprint]{revtex}
\begin{document}
\preprint{hep-th/9401072 \ \ \ IASSNS-HEP-93/87}

\title{On Geometric Entropy}

\author{Curtis Callan\cite{leave} and Frank Wilczek\cite{add}}
\address{
School of Natural Sciences, Institute for
Advanced Study, Princeton, NJ 08540, U.S.A.}
\maketitle

\begin{abstract}We show that a geometrical notion of entropy,
definable in flat space, governs the first quantum correction to
the Bekenstein-Hawking black hole entropy.  We describe two methods
for calculating this entropy -- a straightforward Hamiltonian approach,
and a less direct but more powerful Euclidean (heat kernel) method.
The entropy diverges in quantum field theory in the absence of an
ultraviolet cutoff. Various related finite quantities can be
extracted with further work.  We briefly discuss the corresponding
question in string theory.
\end{abstract}
\narrowtext
\section{Concept of Geometric Entropy}

The microscopic definition $S~=~-{\rm tr} \rho \ln \rho$ of entropy, where
$\rho$ is the density matrix of a quantum-mechanical system, has an appealing
information-theoretic meaning apart from any thermodynamic interpretation.
In quantum field theory, one can define the ``geometric entropy'' associated
with a pure state and a geometrical region by forming the pure state density
matrix, tracing over the field variables inside the region to create an
``impure'' density matrix and then evaluating $S$. In recent years, 't Hooft
\cite{tHooft} and several other authors \cite{Bombelli,Holzhey,Srednicki}
have suggested that quantum-mechanical geometric entropy might be related to
the mysterious Bekenstein-Hawking thermodynamic entropy of a black hole.
In this Letter we will show that geometric entropy is the first quantum
correction to a thermodynamic entropy which reduces to the Bekenstein-Hawking
entropy in the black hole context. It is therefore quite disturbing that,
as first noticed by 't Hooft, this quantity has a divergence that cannot be
renormalized away!

We begin by presenting a slightly unconventional approach to the
Gibbons-Hawking
black hole entropy. The general expression for thermodynamic entropy is
\begin{equation}
S~=~ -(\beta {\partial \over \partial \beta} -1 ) \ln Z
\label{therment}
\end{equation}
where $Z= {\rm tr} e^{-\beta H}$ is the partition function and $H$ is the
generator of time translation invariance. If the underlying
system is a theory of fields $\{\phi_i(\vec x,\tau)\}$, $Z$ can be
computed by doing the Euclidean functional integral over fields
periodic under $\tau\to\tau+\beta$. The periodic identification of the time
coordinate gives a topology to the Euclidean spacetime: $S_1\times R_3$ for
standard thermodynamics and $S_2\times R_2$ for the black hole. In the standard
case, $\beta$ can be given any value and the entropy can be computed for
any temperature. Things are not so simple in the black hole case: The radial
coordinate runs from $r_{\rm horizon}$ to infinity, while $\tau$ is an angle
of period $\beta$. For general $\beta$, there is a two-dimensional conical
curvature singularity at the horizon of deficit angle $\delta= 2\pi
(\beta_H-\beta)/\beta_H$
and the geometry is non-singular ({\it i.e.} $\delta$ vanishes)
only at the Hawking temperature. This is the right physics: a black hole is in
thermodynamic equilibrium only at a specific temperature. Recasting the
general expression for entropy (Eq.\ref{therment}) in terms of deficit angle
rather than $\beta$, one obtains
\begin{equation}
S_{BH}~=~ (2\pi{d \over d \delta} +1 ) \ln Z |_{\delta=0}~.
\label{blackent}
\end{equation}
Since this expression involves derivatives with respect to deficit angle, it
evaluates the Euclidean action on geometries that are not strict solutions of
Einstein's equations, and one might worry that it is ill-defined. However,
Teitelboim\cite{claudio} and Susskind\cite{lenny} have shown that the classical
black hole action can be uniquely extended to conical geometries, thus giving
precise
meaning to Eq.~\ref{blackent}. They have further shown that it generates
  the familiar Bekenstein-Hawking entropy when applied to any of the standard
black hole solutions.

Consider now a superficially quite different quantity: the quantum
entropy of the ground state of a free field, $\phi(\vec x,t)$, generated
by tracing over fields in the $x_1<0$ half space ``inside'' the boundary
surface $x_1=0$. In the basis where the field itself is diagonal, the ground
state wave function is given by the path integral
\begin{equation}
\Psi_0 (\phi(\vec x)) \propto \int {\cal D} {\tilde \phi}
e^{-{1 \over 2}
\int ({\dot {\tilde \phi}}^2 +
(\bigtriangledown {\tilde \phi})^2 +\mu^2 {\tilde \phi}^2) d\vec xd\tau}~.
\label{wavefn}
\end{equation}
The fields $\tilde \phi (\vec x, \tau)$ are defined on all space and all
positive imaginary time, and are subject to the boundary condition
$\tilde \phi (\vec x, 0) = \phi (\vec x)$. The density matrix
$\rho(\phi_+^1 , \phi_+^2)$, where $\phi_+^{1,2}$ are defined for $x_1>0$ only,
is given by a product of two integrals of this type, with an extra trace over a
common boundary value $\phi_-$ on $x_1<0$:
\begin{equation}
\rho(\phi_+^1 , \phi_+^2) = \int{\cal D}\phi_-
\Psi_0(\phi_+^1,\phi_-) \Psi_0(\phi_+^2,\phi_-)
\label{gdstdm}
\end{equation}
If we let the imaginary time in the
two path
integrals run from $0$ to $\pm\infty$, respectively, the density matrix
becomes the path integral over fields defined on the full Euclidean space, with
the $(x_1,\tau)$ plane cut along the $x_1>0$ axis and with $\phi$ taking values
$\phi_+^{1,2}$ above and below the cut. The trace of the density matrix is
obtained by equating the fields across the cut and doing the unrestricted
Euclidean path integral. By the same token, the
trace of the $n$-th power of the density matrix is given by the Euclidean
functional integral over fields defined on an n-sheeted covering of the cut
spacetime. Note that the n-fold cover turns the $(x_1,\tau)$ plane into a flat
cone with deficit angle $2\pi(1-n)$ at the origin.

This geometrical interpretation gives a natural definition of the analytic
continuation of ${\rm tr}\rho^n$ to arbitrary $n$ which in turn allows us to
use
a ``replica trick'' to evaluate the entropy. In order to eliminate any concerns
about normalization, we will construct the entropy out of the explicitly
normalized object $\hat\rho = \rho/{\rm tr}\rho$. Then we can write
\begin{equation}
S_{\rm geom} = -{\rm tr}(\hat\rho {\rm ln} \hat\rho) =
(-{d\over dn} + 1) \ln {\rm tr}\rho^n |_{n=1}~.
\label{replica}
\end{equation}
This expression has the useful property of being independent of overall
rescalings of $\rho$ and we will, from now on, be rather cavalier about the
normalization of $\rho$. The identification of ${\rm tr}\rho^n$ as the
partition
function, $Z_\delta$, on a space of deficit angle $\delta=2\pi(1-n)$ finally
allows us to rewrite Eq.\ref{replica} as
\begin{equation}
S_{\rm geom}~=~(2\pi{d\over d\delta} +1) \ln Z_\delta |_{\delta=0}~.
\label{reptodef}
\end{equation}
But this is formally the same as the expression Eq.\ref{blackent} for the black
hole thermodynamic entropy! To be more precise:  the Bekenstein-Hawking
entropy and the geometrical entropy are the classical and first
quantum contributions to a unified object measuring the response of the
Euclidean path integral to the introduction of a conical singularity in the
underlying geometry.

There is one geometric subtlety to mention: In the free field geometric entropy
problem, the conical geometry obtained by multiply covering flat space is
flat, except for the conical singularity itself.
By contrast, the conical geometry obtained by giving the Euclidean
Schwarzschild solution the wrong $\beta$ has bulk curvature. If the
Bekenstein-Hawking and geometrical quantum entropies are to be just different
orders of approximation to the same thing, the conical geometries must be
{\it exactly} the same. We will achieve this by taking the large mass limit,
in which the Schwarzschild metric goes into Rindler space. In this limit,
curvatures go to zero, the area of the horizon goes to infinity (so that the
quantity of interest becomes entropy per unit horizon area), and the two
conical geometries match precisely. Our theorem, then, is that the
(appropriately defined) geometric entropy of a free field in flat space is just
the quantum correction to the Bekenstein-Hawking entropy of Rindler space.
By introducing curvature in the space on which the free field lives, we could
compute the quantum correction to the entropy of a finite mass black hole.

%
As Holzhey and Srednicki discovered, and we shall further discuss below, the
geometric entropy actually diverges in the absence of an ultraviolet
cutoff. Thus the most important correlations are those between points
just inside and just outside, that is very near to the horizon, where flat
space is an adequate approximation. The essential physics responsible for the
diverging quantum correction to the entropy is not some exotic high curvature
quantum gravity effect, but rather the existence of strong correlations between
nearby field variables in flat space -- a basic tenet of generic {\it
special\/}
relativistic quantum field theories. In projecting the wave functions onto the
ground state as above, we have implicitly assumed
regularity across the horizon.  This means, in the black hole
context, that we have chosen the Unruh vacuum.

\section{Geometric Entropy by Replicas}

We now turn to a more explicit computation of geometric entropy for free
bosonic fields. We will use the replica method directly, in order to give an
independent check of the conical geometry method described above.
By and large, we will be rederiving old (if not widely known) results
\cite{Bombelli}, but the exercise will be an instructive illustration of
the workings of the replica method -- and a verification, in our
context, of its validity.

In Eq.\ref{wavefn} we wrote the ground state wave functional as a Gaussian
functional integral, subject to boundary conditions at
$\tau=0$. Doing the integral by Green's function methods gives the result
\begin{equation}
\Psi_0 (\phi(\bar x)) \propto \exp\{-\int \phi(\bar x)
	\Gamma(\bar x-\bar x^\prime)\phi(\bar x^\prime)d\bar xd\bar x^\prime\}
\end{equation}
where the kernel $\Gamma$ is related to the Dirichlet Green's function
(satisfying $G(\vec x,\tau=0;y)=0$) by
\begin{equation}
\Gamma(\bar x,\bar y) = {1\over 2} {\partial^2 \over \partial \tau ^2}
	G_D(\bar x,\tau;\bar y,0)|_{\tau=0}~.
\label{kernel}
\end{equation}
If we split the base space into two regions and correspondingly split the
field variables into ``inside'' ($\phi_-$) and ``outside'' ($\phi_+$), we can,
in an obvious compressed notation, rewrite the ground state wave function as
\begin{equation}
\Psi_0 \propto \exp\{-(\phi_+~ \phi_-)
\pmatrix{\gamma_{++}&\gamma_{+-}\cr\gamma_{-+}&\gamma_{--}}
		\pmatrix{\phi_+\cr \phi_-}\}
\end{equation}
where $\gamma_{\pm\pm}$ denotes the kernel with arguments restricted to
``inside'' or ``outside'' in various combinations. The integral for the
ground state density matrix, Eq.\ref{gdstdm}, can be done explicitly giving
\begin{equation}
\rho(\phi^1_+; \phi^2_+) \propto
	\exp\{-{1\over 2}(\phi_+^1~ \phi_+^2) \pmatrix{A&2B\cr 2B&A}
	\pmatrix{\phi_+^1\cr \phi_+^2}\}
\label{gaussdm}
\end{equation}
where
\begin{eqnarray}
A &= 2(\gamma_{++} -{1 \over 2} \gamma_{+-} (\gamma_{--})^{-1} \gamma_{-+})\cr
B &= -{1\over2} \gamma_{+-} (\gamma_{--})^{-1} \gamma_{-+}\cr
\label{ABdm}
\end{eqnarray}
where the inverse kernels are inverses of the restricted kernels, {\it not}
restrictions of the full $\Gamma^{-1}$.

Using Eq.\ref{gaussdm}, we can write ${\rm tr}\rho^n$ as a functional integral:
\begin{equation}
{\rm tr}\rho^n =
\int [{\cal D} \phi_+] \exp
 \Biggl\{ -(\phi^1_+ \cdots \phi^n_+) {\cal M}_n
\pmatrix{\phi^1_+\cr \vdots\cr \phi^+_n \cr} \Biggr\}
\end{equation}
where ${\cal M}_n$ is the supermatrix
\begin{equation}
{\cal M}_n = \pmatrix{A & B & 0 & \ldots & 0 & B \cr
B & A & B & \ldots & 0 & 0 \cr
0 & B & A & \ldots & 0 & 0 \cr
\vdots & \vdots & \vdots && \vdots & \vdots \cr
0 & 0 & 0 & \ldots & A & B \cr
B & 0 & 0 & \ldots & B & A \cr}
\end{equation}
and $A$ and $B$ are the kernels defined in Eq.\ref{ABdm}. Doing the Gaussian
functional integral gives the result ${\rm tr}\rho^n =1/\sqrt{{\rm det}{\cal
M}_
 n}$.
The problem is to compute this functional determinant as an explicit
function of $n$.

According to the discussion following Eq.\ref{replica}, we can freely rescale
$\rho$ without affecting the value of the entropy. Let us define a new matrix
\begin{equation}
\widetilde{\cal M}_n =
\pmatrix{2 & -C & 0 & \ldots & 0& -C\cr
-C & 2 & -C & \ldots & 0 & 0\cr
 0 & -C & 2 & \ldots & 0 & 0\cr
\vdots & \vdots & \vdots &&\vdots & \vdots\cr
0 & 0 & 0 & \ldots & 2 & -C\cr
-C & 0 & 0 & \ldots & -C & 2\cr}
\end{equation}
with $C=-2A^{-1}B$. $C$ can be more symmetrically expressed as
$C = (2-D)^{-1}D$ where
\begin{equation}
D \equiv \gamma_{++}^{-1}\gamma_{+-} \gamma_{--}^{-1}\gamma_{-+}
\label{entmat}
\end{equation}
is a symmetric matrix acting on the space of $\phi_+$ variables.
Since ${\rm det}\widetilde{\cal M}_n =
2^n {{\rm det}{\cal M}_n / ({\rm det} A)^n}$, replacing ${\cal M}_n$ by
$\widetilde{\cal M}_n$ amounts to rescaling $\rho$ by a factor of
${1\over 2}{\rm det} A$
and has no effect on the entropy.

If the operator $C$ were a c-number, we could explicitly diagonalize
$\widetilde{\cal M}_n$ and show that
\begin{equation}
\det {\widetilde{\cal M}_n} =\prod_{r=1}^n\{2-2~C \cos({2\pi r\over n})\}
		= 2^n{(1 - \xi^n)^2 \over (1+\xi^2)^n}
\end{equation}
where $\xi(C)$ is defined by $C = 2\xi /( \xi^2 +1)$.
Now $C$ is an operator, not a number, but we may diagonalize it
and apply the above reasoning eigenvalue by eigenvalue:
\begin{equation}
\det {\widetilde{\cal M}_n} = 2^n\prod_{C_i}{(1 - \xi(C_i)^n)^2 \over
		(1+\xi(C_i)^2)^n}~.
\end{equation}
For the usual reasons, we may drop the factors of $2^n(1-\xi^2)^n$ since
they have no effect on the entropy. The operations called for by the replica
trick (Eq.\ref{replica}) then give the entropy as
a {\it sum} over contributions from each eigenvalue of $C$:
\begin{equation}
S_{\rm replica} = -\sum_{C_i}\{{\rm ln}(1-\xi(C_i)) +
{\xi(C_i)\over 1-\xi(C_i)} {\rm ln}\xi(C_i)\}~.
\label{eigsum}
\end{equation}
Although it is not obvious, the eigenvalues of $C$ all lie in $(0,1)$,
as do the values of $\xi(C)$, so the sum for $S$ is positive term-by-term.

The essential point of all this is that the geometric entropy is calculable
from the spectrum of a kernel $C$, acting on the ``outside'' space. Actually,
$C$ is constructed out of the standard Green's functions in a very awkward
way, and it is better to look at the spectrum of a more convenient function of
$C$. The operator $D$, defined in Eq.\ref{entmat}, and related to $C$ by
$C=(2-D)^{-1}D$, is one possibility. It turns out that the operator
$E=(1-D)^{-1}D$ is much nicer: it can be expressed as a single convolution
\begin{equation}
E \equiv -(\Gamma^{-1})_{+-}~\Gamma_{-+}
\label{newker}
\end{equation}
of the original kernel $\Gamma$,
defined in Eq.\ref{kernel}, with its inverse (as usual, the subscripts
indicate restrictions of the arguments to ``inside'' or ``outside'' values).
The eigenvalues of $E$ are positive and the relation to $\xi$ is
\begin{equation}
\xi(E) = 1+ {2\over E}(1-\sqrt{1+E})~.
\end{equation}
The entropy sum, Eq.\ref{eigsum}, can of course be expressed directly in terms
of the eigenvalues $E$:
\begin{equation}
S_{\rm replica} = \sum_{E_i}
\{{\rm \ln}\sqrt{{E_i\over 2}}+\sqrt{1+E_i}
{\rm \ln}({1\over\sqrt{E_i}}+\sqrt{1+
{1\over E_i}})  \}
\label{entsum}
\end{equation}

\section{Explicit Calculations}

Consider a one-dimensional free field of mass $\mu$,
confined to a box $-L<x<L$ to eliminate infrared problems.
We want to compute the entropy of the ground
state density matrix obtained by tracing over the fields in $-L<x<0$
(so that ``inside'' (``outside'') correspond to $x<0$ ($x>0)$).
We have outlined two methods, Eq.\ref{eigsum} and Eq.\ref{reptodef}, which
we might call Hamiltonian and Euclidean, respectively. They should agree with
each other and with conformal field theory, where applicable \cite{Holzhey}.

Let us first take the Hamiltonian approach and try to compute the eigenvalue
sum in Eq.\ref{eigsum}. The most straightforward approach is to discretize
the system and then solve the resulting finite-dimensional eigenvalue problem
numerically. Although this calculation is instructive, and we have done it
to double-check certain things, it turns out to be hard to extract the
continuum-limit behavior of $S$ from it. Let us instead try an approximate
analytic approach. In the absence of the infrared cutoff ($L\to\infty$),
the kernels needed to construct $E$ are explicit and simple:
\begin{eqnarray}
\Gamma(x,y)^{-1} =&K_0(\mu(x-y))/\pi\cr
\Gamma(x,y) =&\mu K_1(\mu(x-y))/(\pi (x-y))
\end{eqnarray}
The eigenvalue problem to be solved is then (this eigenvalue problem appears
in \cite{Bombelli}; our contribution is only to show its connection, via the
replica trick, with the black hole entropy problem)
\begin{equation}
\int_0^\infty dy\int_0^\infty dz
	K_0(\mu(x+z)) {K_1(\mu(z+y))\over\pi^2\mu^{-1}(z+y)} \psi(y) = E\psi(x)
\end{equation}
where $x>0$. It becomes especially simple in the $\mu\to 0$ limit:
\begin{equation}
-\int_0^\infty dy\int_0^\infty dz{{\rm \ln}(x+z)\over \pi^2(y+z)^2}
                \psi(y) = E \psi(x).
\end{equation}
One can easily show by contour integration that
$\psi(x) = \exp(\imath\omega {\rm \ln} x)$ is an eigensolution with eigenvalue
$E ={\rm sinh}^{-2}(\pi\omega)$. The log-periodic nature of these
eigenvectors is very reminiscent of the Unruh basis for the calculation of
the Hawking radiation.

To calculate the entropy, we must discretize the spectrum. This is normally
done by imposing boundary conditions. A Dirichlet
boundary condition at some large $x=L$ is needed, in any case, to regulate
infrared divergences. It seems natural, in addition, to impose
an ultraviolet cutoff via a further Dirichlet condition at
some small $x=\epsilon$ (this is just a reasonable guess: the integral
equation picks its own boundary condition!). The eigensolutions are then
\begin{eqnarray}
\psi_E(x) =& \sin(\omega(E) {\rm ln}(x/\epsilon))\cr
\omega(E_n)&{\rm ln}(L/\epsilon) = \pi n ~,
\end{eqnarray}
where $\omega(E)=\pi^{-1}\sinh^{-1}(\sqrt{E})$ is the eigenvalue
function. It also seems reasonable to keep a finite number of degrees of
freedom by cutting off the mode number $n$ at $L/\epsilon$. The picture so
obtained agrees very nicely with the numerical work we have done on the
discretized problem. The crucial point is that, in the limit
$L/\epsilon\to\infty$, the eigenvalues become dense with an eigenvalue density
\begin{equation}
{d\over dE} \rho(E) =
    -{\omega^\prime(E)\over\pi}{\rm ln}(L/\epsilon)
\end{equation}
which diverges logarithmically with the cutoff.

Note that the density of
states per unit $\omega$ interval is constant, suggesting that it might
make more sense to rewrite the continuum limit of the entropy expression,
Eq.\ref{entsum}, as an integral over $\omega$, expressing the integrand as
a function of $\omega$ rather than $E$. The result of doing this is
\begin{equation}
S_{{\rm replica}} = {{\rm ln}({L\over\epsilon})\over\pi}\int_0^\infty d\omega
\{
{2\pi\omega\over(e^{2\pi\omega}-1)}-{\rm ln}(1-e^{-2\pi\omega})\},
\label{rindent}
\end{equation}
an expression which is easily recognized as the entropy of a one-dimensional
gas of massless bosons
confined to a box of size ${\rm ln}({L\over\epsilon})$ at a temperature
$\beta=2\pi$.
The integral can be done exactly (see any statistical mechanics text) and
yields
\begin{equation}
	S_{\rm replica}\sim {1\over 6} {\rm ln}(L/\epsilon)~+~O(L^0).
\end{equation}
The essential fact is that the entropy {\it diverges}, not because the
contribution of large or small eigenvalues blows up, but rather because the
eigenvalue {\it density} blows up everywhere like ${\rm ln}(L/\epsilon)$. This
is a direct reflection of the Unruh-style log-periodic behavior of the
eigenvectors.

Some further remarks are in order. The formula in Eq.\ref{rindent} is precisely
the expression derived by Kabat and Strassler \cite{kabat} for the black hole
entropy in the large-mass (Rindler space) limit. This is another piece of
evidence for the identity of geometric and black hole entropy. Also,
Eq.\ref{rindent} cries out to be generalized to fermions: the corresponding
formula for the entropy of massless one-dimensional fermions at zero chemical
potential (to pick out the vacuum state) is obtained by a strategic sign
change reflecting Fermi-Dirac as opposed to Bose-Einstein statistics:
\begin{equation}
S_{\rm replica} = {{\rm ln}({L\over\epsilon})\over\pi}
\int_0^\infty d\omega \{
{2\pi\omega\over(e^{2\pi\omega}+1)}+{\rm ln}(1+e^{-2\pi\omega})\}.
\label{rindferment}
\end{equation}
This integral can also be done exactly, giving the result
\begin{equation}
	S_{\rm replica}^{\rm Fermi}\sim {1\over 12} {\rm ln}(L/\epsilon)~+~O(L^0).
\end{equation}
It thus seems likely that the divergence of the geometric entropy will not
be cancelled by any supersymmetric Bose-Fermi cancellation. A direct
calculation
of the fermionic entropy would be interesting to do.

Next, let us use heat kernel methods to compute the functional determinant
appearing in the Euclidean expression, Eq.\ref{reptodef}, for the entropy.
This calculation can easily be done in arbitrary dimension: we study
a Euclidean field theory on the space $C_\delta\times M_{D-2}$ where
$C_\delta$ is a two-dimensional cone of radius $L$, deficit angle
$\delta$ and $M_{D-2}$ is a flat ($D-2$)-dimensional transverse space with
total volume $V_{D-2}$. For a free massive scalar field, the path integral is
determined by the functional determinant of the kinetic energy operator:
\begin{equation}
{\rm ln} Z_\delta = -{1\over 2}\ln {\rm det}(-\Delta+\mu^2)~.
\end{equation}
To calculate the determinant, we use the eigenvalues $\lambda_n$ of
$-\Delta$ on the manifold under study to define the ``heat kernel''
\begin{equation}
\zeta(t)= {\rm tr}(e^{t\Delta})=\sum_n e^{-t\lambda_n}.
\end{equation}
$\zeta$ depends implicitly on the deficit angle $\delta$ of the geometry.
A cut-off version of the determinant can then be defined as
\begin{equation}
{\rm ln~det~}(-\Delta+\mu^2)=
-\int_{\epsilon^2}^\infty{dt\over t}\zeta(t)e^{-{\mu}^2 t}
\label{heatdet}
\end{equation}
where $\epsilon$ is a short distance cutoff with dimensions of length.
The heat kernel expression for the entropy is then
\begin{equation}
S = {1\over 2}\int_{\epsilon^2}^\infty{dt\over t}
(2\pi{d\over d\delta}+1) \zeta(t) e^{-{\mu}^2 t}|_{\delta=0}
\label{heatent}
\end{equation}

The utility of the heat kernel stems from the fact that its expansion
in powers of $t$ near $t=0$ is governed by simple geometrical properties
of the manifold\cite{balian,alvarez}. On the two-dimensional manifold
$C_\delta$, for example,
\begin{equation}
\zeta_2(t) = t^{-1}{(2\pi-\delta)\over 8\pi} L^2 +
	{1\over {12}} ({2\pi \over \beta} - {\beta \over 2\pi})
+ O(t/L^2) ~.
\end{equation}
On the full manifold $C_\theta\times M_{D-2}$, there is an extra factor
$V_{D-2}/(4\pi t)^{{D-2\over 2}}$ coming from the trivial Laplacian on the
transverse space. Substituting this into Eq.\ref{heatent}, we get an
expansion of the entropy in powers of $L^{-1}$:
\begin{eqnarray}
S = &{V_{D-2}\over(2\sqrt{\pi})^{D-2}}\int_{\epsilon^2}^\infty {dt\over
t^{D/2}}
	e^{-\mu^2 t}({1\over 12}+O(t/L^{-2}))\cr
& \cr
    = &(1/6){\rm ln}(1/\epsilon) + \ldots~~(D=2)\cr
& \cr
    = & {V_{D-2}\over(2\sqrt{\pi})^{D-2}(D/2-1)}\epsilon^{2-D}+\ldots~~(D>2)~.
\end{eqnarray}

Note that the replica trick differentiation with respect to deficit
angle annihilates the $t^{-D/2}$ term in $\zeta(t)$. This would have given
an extensive entropy proportional to the total volume of the $D$-manifold.
The surviving $t^{0}$ term gives a boundary entropy proportional to the volume
of the boundary manifold $M_{D-2}$. It is logarithmically divergent in $D=2$
and correspondingly more divergent in higher dimensions. The divergent piece in
$D=2$ agrees with our previous ``direct'' calculation and with \cite{Holzhey}
while the quadratically divergent boundary area entropy in
$D=4$ agrees approximately with
 Srednicki's numerical evaluation\cite{Srednicki}. In fact it is
unlikely that the coefficient of the leading divergence is
universal -- that is, independent of the regulator
procedure -- for $D \not= 2$, since it depends on the scale of
$\epsilon$.
Further terms in the expansion of $\zeta(t)$ give
finite volume corrections which vanish in the limit $L\to\infty$.

\section{Comments}

1. There are also finite quantities lurking within this circle of ideas which
seem likely to be of physical significance.
One class of such quantities is the difference between
geometric entropy for the same region but with respect to different
states (for example, thermal or solitonic states). Another class is
the geometric entropy for the ground state in its dependence on the
shape of the bounding surface. Specifically, we suspect that one
can formulate an expansion for the geometric entropy associated with
a bounding surface similar to an effective action, where
the coefficients of invariants
associated with high dimension operators ({\it i.e.} powers of curvatures
and their derivatives) will be cutoff independent.
Another arises in the dependence on the geometry of the embedding
space, for a given bounding surface.
Yet another occurs in topological field theories, where the
geometric entropy associated with excision of a topologically
non-trivial subspace defines a finite topological invariant.
The heat kernel algorithm appears to provide a practical approach to such
questions.

2. String theory is thought to provide a consistent framework for
calculating quantum corrections to classical general relativity,
and in particular to be free of the ultraviolet
divergences that plague standard local field theories in this regard.
It is therefore of great interest to consider quantum corrections to
the Bekenstein-Hawking entropy in the context of string theory
\cite{susskind}.
While a complete solution to this problem is far out of sight, our
considerations
here suggest a subproblem that appears to us both central and tractable.
As we have seen, there is a divergent piece of the entropy arising in
order $g^0$ (free field theory) that can be identified even from the
partition function in flat space,
or
more precisely in flat space with a weak conical singularity.  In future
work we will attempt to
get at this by studying orbifolds \cite{orbi} formed from flat space
by identifications under discrete rotation groups $C_n$, by analytic
continuation in $n$.

3. How should one react to the divergence of the first quantum correction
to the Bekenstein-Hawking entropy?  Perhaps one's first instinct is
to attempt to subtract or renormalize it away.  But subtraction
is impossible because of the non-trivial dependence of the correction
on area; while renormalization, even if possible, would leave the
numerical value of the entropy indeterminate.  In any case
we think that our physical interpretation of the corrections as reflecting
concrete measurable correlations across the horizon precludes any
such procedure.  Thus we are forced, within quantum field theory,
either to abandon the equilibrium formula $dM ~=~ T dS$ or to
let the temperature vanish.  In string theory the correction to the
Hawking result may be finite, but even then it need not be small numerically.

Acknowledgement:
We want to thank L. Susskind for extensive discussions, and for informing
us of the conical singularity approach to black hole entropy.
We also wish to thank F. Larsen and S. Shenker for helpful,
stimulating discussions.  We also wish to thank S. Dowker for pointing
out a mistake in a previous version of Eq. (35).

\end{document}